%% file: main.tex
\documentclass{article}

\PassOptionsToPackage{numbers,sort&compress}{natbib}

\usepackage[preprint]{neurips_2026}

\usepackage[utf8]{inputenc}
\usepackage[T1]{fontenc}
\usepackage{hyperref}
\usepackage{url}
\usepackage{booktabs}
\usepackage{amsfonts}
\usepackage{nicefrac}
\usepackage{microtype}
\usepackage{xcolor}

\usepackage{graphicx}
\usepackage{listings}
\usepackage{float}
\usepackage{xspace}
\usepackage{caption}
\usepackage{wrapfig}
\usepackage{xurl}
\usepackage{pifont}
\usepackage{fontawesome5}

\newcommand{\SkCC}{%
    \textsc{SkCC}\xspace
}
\newcommand{\SkIR}{%
    \textsc{SkIR}\xspace
}

\lstdefinelanguage{json}{
  basicstyle=\ttfamily\small,
  numbers=left,
  numberstyle=\tiny\color{gray},
  stepnumber=1,
  numbersep=8pt,
  showstringspaces=false,
  breaklines=true,
  frame=single,
  frameround={t}{t}{t}{t},
  rulecolor=\color{black!30},
  backgroundcolor=\color{black!3},
  keywordstyle=\color{blue!60!black}\bfseries,
  stringstyle=\color{teal!50!black},
  commentstyle=\color{gray},
  morestring=[b]",
  morecomment=[l]{//},
  morecomment=[s]{/*}{*/},
  literate=
    *{0}{{{\color{orange!70!black}0}}}1
     {1}{{{\color{orange!70!black}1}}}1
     {2}{{{\color{orange!70!black}2}}}1
     {3}{{{\color{orange!70!black}3}}}1
     {4}{{{\color{orange!70!black}4}}}1
     {5}{{{\color{orange!70!black}5}}}1
     {6}{{{\color{orange!70!black}6}}}1
     {7}{{{\color{orange!70!black}7}}}1
     {8}{{{\color{orange!70!black}8}}}1
     {9}{{{\color{orange!70!black}9}}}1,
}

\lstdefinestyle{plaintext}{
  basicstyle=\ttfamily\small,
  numbers=left,
  numberstyle=\tiny\color{gray},
  showstringspaces=false,
  breaklines=true,
  frame=single,
  frameround={t}{t}{t}{t},
  rulecolor=\color{black!30},
  backgroundcolor=\color{black!3},
}

\title{\SkCC{}: Portable and Secure \underline{Sk}ill \underline{C}ompilation for \underline{C}ross-Framework LLM Agents}

\author{%
  Yipeng Ouyang \\
  Sun Yat-sen University \\
  \texttt{ouyyp5@mail2.sysu.edu.cn} \\
  \And
  Yi Xiao \\
  Sun Yat-sen University \\
  \texttt{xiaoy398@mail2.sysu.edu.cn} \\
  \And
  Yuhao Gu \\
  Sun Yat-sen University \\
  \texttt{guyh9@mail2.sysu.edu.cn} \\
  \And
  Xianwei Zhang \\
  Sun Yat-sen University \\
  \texttt{zhangxw79@mail.sysu.edu.cn} \\
}

\makeatletter
\renewcommand{\@noticestring}{%
  \rule{0.35\textwidth}{0.4pt}\\[2pt]
  \begin{minipage}{0.92\textwidth}
  Preprint. Project initiated in March 2026; Core compiler repository open-sourced via GitHub on April 3rd, 2026; Initial draft completed in April 2026; Paper submitted in May 2026.%
  \end{minipage}
}
\makeatother

\begin{document}

\maketitle


\begin{center}
\vspace{-8pt}
\faGithub \quad \url{https://github.com/Nexa-Language/Skill-Compiler} \\
\includegraphics[height=0.85em]{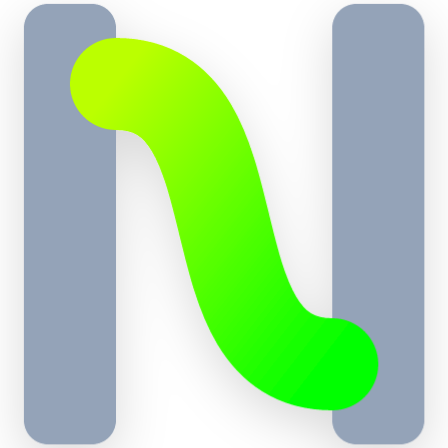} \quad \url{https://skcc.nexa-lang.com/}
\vspace{4pt}
\end{center}


\input{sections/00_abstract}

\input{sections/01_introduction}

\input{sections/02_background}

\input{sections/03_design}

\input{sections/04_evaluation}

\input{sections/05_conclusion}


\begin{ack}
This work is a preprint currently under review. The \SkCC{} project (originally named Nexa-Skill-Compiler, NSC) initiated in March 2026; Core compiler repository open-sourced via GitHub (\url{https://github.com/Nexa-Language/Skill-Compiler}) on April 3rd, 2026; Initial draft completed in April 2026; Paper submitted in May 2026. We document this timeline to clarify the independent nature of our contributions relative to concurrent developments in the agent skill ecosystem.

This project is part of the \href{https://www.nexa-lang.com/}{Nexa-lang Project}. The homepage for \SkCC{} is available at \url{https://skcc.nexa-lang.com/}. All rights reserved by \href{https://www.nexa-lang.com/}{Nexa-lang Project} \& \href{https://arcsysu.tech/}{arcSYSu Lab, Sun Yat-sen University}.
\end{ack}


\bibliographystyle{unsrtnat}
\bibliography{reference}


\appendix

\input{sections/07_appendix}

\end{document}

%% file: sections/00_abstract.tex

\begin{abstract}
LLM agents increasingly rely on reusable skills (e.g., SKILL.md) to execute complex tasks, yet these artifacts lack portability: agent frameworks are highly sensitive to prompt formatting, leading to a large performance variation for the same skill. Nevertheless, most skills are authored once as format-agnostic Markdown, necessitating costly per-framework rewrites and also leaving security largely unaddressed, with widespread vulnerabilities in practice.
To address this, we present \SkCC{}, a compiler for LLM agents that introduces classical compilation design into agent skill development. \SkCC{} centers on \SkIR{}, a strongly-typed intermediate representation that decouples skill semantics from framework-specific formatting, thus enabling portable deployment across agent frameworks. Atop of this IR, a static Optimizer enforces security constraints, blocking vulnerabilities before deployment. Implemented as a four-phase pipeline, \SkCC{} effectively reduces adaptation complexity from $O(m \times n)$ to $O(m + n)$ across $m$ skills and $n$ frameworks. Experiments on SkillsBench demonstrate that \SkCC{} delivers consistent and substantial gains over original counterparts, with pass rate increases from 21.1\% to 33.3\% on Claude Code and from 35.1\% to 48.7\% on Kimi CLI. Further, the design achieves sub-10ms compilation latency, 94.8\% proactive security trigger rate, and 10--46\% runtime token savings across frameworks.
\end{abstract}

%% file: sections/01_introduction.tex

\section{Introduction}

The rapid advancement of large language models (LLMs) has catalyzed a new generation of autonomous agent systems~\cite{wooldridge1995,yao2023,wang2024survey}. Agent frameworks such as Anthropic Claude Code~\cite{claude_code}, OpenAI Codex~\cite{codex_docs}, Google Gemini CLI~\cite{gemini_docs}, and Kimi CLI~\cite{kimi_docs} provide terminal-based agent environments where LLMs interact with tools, file systems, and external services. Skills, structured prompt artifacts following the SKILL.md specification~\cite{agentskills_spec}, have emerged as the {\it de facto} standard for encoding domain-specific knowledge, employing progressive disclosure~\cite{xu2026} that loads lightweight metadata at initialization and retrieves full content on demand. As the ecosystem matures, the number of community-contributed skills has grown rapidly, with repositories such as Anthropic-skills~\cite{anthropic_skills}, ecc-skills~\cite{everything_claude_code}, and sentry-skills~\cite{sentry_skills} collectively hosting thousands of reusable skill artifacts.

However, a growing body of evidence reveals that LLM performance is highly sensitive to the structural format in which skills are presented~\cite{he2024}. For example, Claude performs substantially better when skills use XML semantic layering~\cite{claude_prompting}, GPT-series models benefit from XML-tagged Markdown that avoids the "format tax" of JSON~\cite{openai_structured}, and deeply nested data is parsed most accurately in YAML~\cite{improvingagents2025}. Yet the current ecosystem assumes format-agnostic delivery: the same SKILL.md is deployed identically across all frameworks, ignoring these well-documented format preferences. The same skill can exhibit a large performance variation depending solely on how it is formatted for a given model. Beyond format compatibility, the skill ecosystem faces an equally pressing security challenge. Snyk's audit~\cite{snyk2026} found that over one third community skills contain security vulnerabilities, including many confirmed malicious payloads. The SKILL.md specification acknowledges the need for negative boundaries~\cite{agentskills_explained}, yet most existing skills lack such constraints, and no systematic mechanism exists to enforce security properties before skills reach an agent's context window. These two challenges, format sensitivity and security vulnerability, are not independent. They both stem from the fundamental assumption that a single, static Markdown file can serve all frameworks and all threat models simultaneously.

We present \SkCC{}, a systematic skill compilation design that addresses both the portability and security challenges of cross-framework skill deployment. The central insight is that a unified intermediate representation, \SkIR{}, can decouple skill authoring from framework-specific formatting, enabling each skill to be written once and compiled to multiple frameworks. This mirrors the classical compiler architecture that just as LLVM IR enabled a single frontend to target diverse hardware backends, \SkIR{} enables a single skill source to target diverse agent frameworks. \SkCC{} operates through a four-phase pipeline: \ding{172}a Syntax Parser extracts AST from raw SKILL.md, \ding{173}an IR Builder transforms it into a strongly-typed \SkIR{}, \ding{174}a Security Optimizer enforces safety constraints via Anti-Skill Injection, and \ding{175}a polymorphic Target Emitter renders the validated IR into framework-native formats. This architecture reduces the adaptation complexity from $O(m \times n)$ to $O(m + n)$.

\begin{figure}[t]
  \centering
  \includegraphics[width=0.95\textwidth]{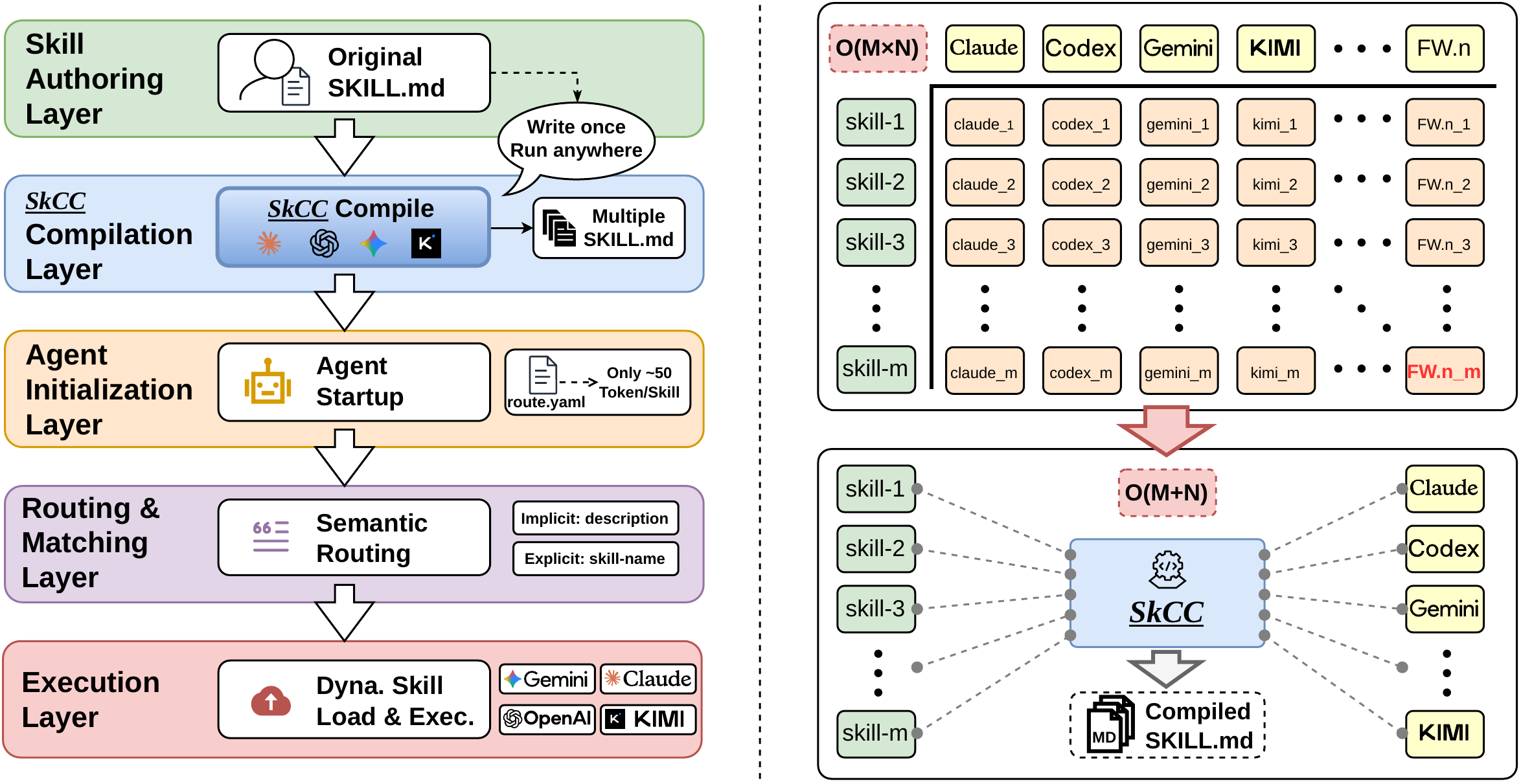}
  \caption{Left: Agent workflow with \SkCC{} integration. Skills are authored once as SKILL.md, compiled to framework-native formats, and loaded via progressive routing manifests at agent initialization. Right: Adaptation complexity reduction from $O(m \times n)$ to $O(m + n)$. Traditional per-framework rewriting requires $m \times n$ manual adaptations. \SkCC{} decouples skills and frameworks through a unified IR, requiring only $m$ skill sources and $n$ Emitter implementations.}
  \label{fig:complexity_and_flow}
\end{figure}

Our key contributions are as follows:

\begin{itemize}
  \item \textbf{We identify a structural gap in the agent skill ecosystem}: format sensitivity is a first-class concern in skill deployment, and the growing diversity of agent frameworks makes manual per-framework adaptation infeasible, motivating a compiler-based solution with a unified intermediate representation.

  \item \textbf{We propose \SkCC{}}, a four-phase skill compilation design that achieves portable deployment via \SkIR{}, and secure execution through Anti-Skill Injection and semantic validation. By introducing a unified IR and a polymorphic emission layer, \SkCC{} decouples skill authoring from framework-specific formatting, reducing the $O(m \times n)$ adaptation burden to $O(m + n)$ while enforcing security constraints before deployment.

  \item \textbf{We implement and evaluate \SkCC{} across mainstream agent frameworks}, demonstrating consistent pass rate improvements (up to +13.5\%), sub-10ms compilation latency, 94.8\% Anti-Skill Injection coverage, and 10--46\% runtime token savings, demonstrating strong gains in portability, security, and efficiency across frameworks.
\end{itemize}

%% file: sections/02_background.tex

\section{Background and Motivation}
\label{sec:bg_motivation}

\subsection{Background}
\label{sec:bg}

Our Design covers two areas: how agent skills are structured and used in practice, and the classical compilation principles that inform our approach.

\textit{Agent Skills: Structure and Usage.}
Modern LLM agent systems~\cite{wooldridge1995,yao2023,wang2024survey} execute complex tasks by composing tool calls, file system operations, and external service interactions. To encode domain-specific knowledge in a reusable form, the community has converged on SKILL.md~\cite{agentskills_spec}, a portable specification consisting of YAML frontmatter for metadata and a Markdown body for executable instructions. Skills are loaded through progressive disclosure~\cite{xu2026}: a lightweight routing manifest ($\sim$50 tokens per skill) is loaded at initialization, and full content is retrieved on demand when semantically matched to the user's task. Skills interact with external systems through the Model Context Protocol (MCP)~\cite{mcp_spec}, a standardized interface for connecting agents to tools and services. Agent skills and their MCP interactions rely on several structured data formats: XML (tag-delimited trees), JSON (key-value pairs), YAML (indentation-based nesting, superior LLM parsing accuracy for deeply nested structures~\cite{improvingagents2025}), and Markdown (lightweight markup). These syntactic differences directly affect how accurately LLMs extract and follow instructions.

\textit{Classical Compilation Principles.}
A traditional compiler~\cite{aho1986,muchnick1997} transforms source code through a multi-phase pipeline: lexical analysis, parsing into an AST, semantic analysis, IR generation, optimization, and target code generation. The critical architectural insight is the role of the IR: by introducing a unified intermediate layer, compilers decouple frontend language parsing from backend code generation, reducing the $m \times n$ support problem to $m + n$~\cite{strong1958,lattner2004,lattner2021mlir}. Security optimization at compile time, such as stack canary insertion and bounds checking~\cite{szekeres2013}, further demonstrates that compilers can enforce safety properties before code executes.

\subsection{Related Work and Challenges}
\label{sec:rw}

Having established the foundational concepts, we now analyze recent work and identify limitations that motivate our approach.

\textit{Format Sensitivity and Skill Retrieval.}
LLM performance is highly sensitive to prompt formatting, with up to 40\% variation from format changes alone~\cite{he2024}. Framework-specific preferences are well-documented: Claude benefits from XML semantic layering~\cite{claude_prompting,reddit_claudeai,philip2025}, GPT-series models suffer from a ``format tax'' with JSON~\cite{openai_structured,kinney2026}, and YAML achieves superior parsing accuracy for nested data~\cite{improvingagents2025}. CFPO~\cite{liu2025} jointly optimizes content and format through iterative refinement, but its search-based approach is computationally expensive and produces instance-specific rather than reusable rules. On the retrieval side, recent work explores generation, augmentation, graph, and embedding skill retrieval~\cite{wang2024toolgen,su2026,edge2024,reimers2019}, and Liu et al.~\cite{liu2026wild} show that query-specific refinement yields modest post-retrieval gains. These works share a common assumption, that skills once retrieved are format-agnostic and require no structural adaptation.

\textit{Compilation and Security for Agent Skills.}
Applying compilation techniques to LLM systems has gained traction: Mikek et al.~\cite{mikek2026} demonstrate compiler-LLM cooperation for agentic code optimization, and Kim et al.~\cite{kim2023} use compiler orchestration for parallel function calling. SkVM~\cite{chen2026} also explores compilation concepts for agent skills with a JVM-like architecture supporting capability profiling and AOT/JIT optimization. On the security dimension, Snyk's audit~\cite{snyk2026} finds 37\% of 3,984 community skills contain vulnerabilities, yet the SKILL.md specification's recommended negative boundaries~\cite{agentskills_explained} are rarely followed~\cite{kumar2026}, and recent work on secure code generation~\cite{wang2026secpi} operates at the code rather than skill level.

\textit{Challenges.}
The preceding analysis reveals a structural gap: existing systems either ignore format sensitivity, address it through expensive instance-specific search, or focus on semantic capability without format-syntax adaptation, while, to our knowledge, no system provides systematic compile-time security enforcement for agent skills (Table~\ref{tab:qualitative_comparison}).

\begin{table}[t]
  \centering
  \caption{Capability Coverage of Related Systems}
  \label{tab:qualitative_comparison}
  \small
  \begin{tabular}{lcccc}
    \toprule
    \textbf{Method} & \textbf{Format Adapt.} & \textbf{Security} & \textbf{Multi-Framework} & \textbf{Complexity} \\
    \midrule
    SkVM~\cite{chen2026} & Semantic only & \texttimes & \checkmark & $O(m \times n)$ \\
    CFPO~\cite{liu2025} & Iterative & \texttimes & \texttimes & $O(k \times m)$ \\
    Wild Retrieval~\cite{liu2026wild} & Query-specific & \texttimes & \checkmark & $O(m \times n)$ \\
    \textbf{Desired} & \textbf{IR-driven} & \textbf{\checkmark} & \textbf{\checkmark} & \textbf{$O(m + n)$} \\
    \bottomrule
  \end{tabular}
\end{table}

These challenges share a common architectural root. Supporting diverse skills across diverse frameworks requires a decoupling layer, a unified intermediate representation that separates skill semantics from framework-specific formatting, combined with compile-time analysis that enforces security constraints before deployment. Rather than treating skills as static text files that must be manually rewritten for each target, a compiler-based methodology treats them as compilable artifacts: authored once in a canonical form, analyzed and optimized at compile time, and emitted into framework-native formats through platform-specific backends. This separation of concerns mirrors the classical compiler architecture that revolutionized systems programming, and we argue it is equally necessary for the emerging agent skill ecosystem.

%% file: sections/03_design.tex

\section{\SkCC{} Design}
\label{sec:design}

\begin{figure}[t]
  \centering
  \includegraphics[width=\textwidth]{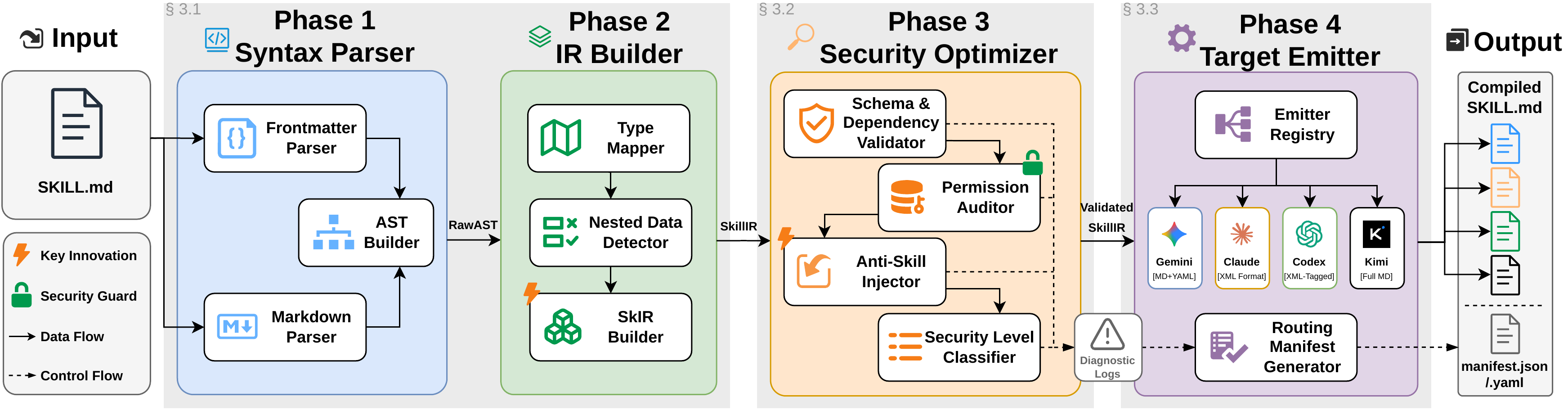}
  \caption{\SkCC{}'s four-phase compilation pipeline. A unified SKILL.md source is parsed into a raw AST (Syntax Parser), transformed into a strongly-typed \SkIR{} (IR Builder), validated and optimized by compile-time security analysis (Security Optimizer), and emitted into framework-native formats (Target Emitters).}
  \label{fig:architecture}
\end{figure}

\SkCC{} is a compilation pipeline that accepts a single SKILL.md source and produces framework-native skill artifacts through four phases (Figure~\ref{fig:architecture}). Phases 1--2 (Syntax Parser and IR Builder, \S\ref{sec:parser}) extract structured, typed representations from raw Markdown, producing \SkIR{}, a unified intermediate representation that decouples skill semantics from framework-specific formatting. Phase~3 (Security Optimizer, \S\ref{sec:optimizer}) optimizes the IR through a chain of compile-time analyses that validate structure, audit permissions, inject safety constraints, and assign security levels. Phase~4 (Target Emitter, \S\ref{sec:emitter}) renders the optimized IR into framework-native formats through a polymorphic emission layer. The critical architectural property is that Phases 1--3 execute once per skill; the resulting optimized \SkIR{} is then shared across all emission targets, reducing the adaptation complexity from $O(m \times n)$ to $O(m + n)$.

\subsection{Syntax Parsing and IR Building}
\label{sec:parser}

\paragraph{Syntax Parser.}
Raw SKILL.md files interleave structural metadata (YAML frontmatter) with free-form instructional text (Markdown body), creating ambiguity for downstream consumers. The Syntax Parser eliminates this ambiguity by aggressively separating concerns at the syntactic level: metadata is deserialized into a typed routing table, while the Markdown body is lowered into a deterministic abstract syntax tree where procedure steps, code blocks, and examples are explicitly classified. This separation ensures that every subsequent phase operates on structured, unambiguous data rather than raw text, and it enables the compiler to reason about skill structure independently of authoring style.

\paragraph{IR Builder.}
The IR Builder transforms the raw AST into \SkIR{}, a strongly-typed intermediate representation. The key methodological decision is to normalize heterogeneous skill content into a uniform, typed structure that captures what a skill \emph{means} independently of how it is \emph{formatted}. Rather than preserving Markdown-level details, \SkIR{} abstracts skill information into semantic categories (procedures, permissions, schemas, constraints), each with well-defined types and validation rules. This abstraction serves two purposes. First, it provides a single source of truth that all downstream phases can consume without re-parsing or re-interpreting the original text. Second, it creates a clean boundary between skill authoring and skill deployment: authors write in a single canonical format, and the IR insulates them from the formatting requirements of individual frameworks. A concrete \SkIR{} instance is provided in Appendix~\ref{app:skillir_example}.

A representative capability of the IR level is nested data detection: when a skill declares schemas with nesting depth exceeding a threshold, the IR records a flag that downstream Target Emitters consult to decide whether to render structured data in a format suited for deep nesting. This illustrates the IR's broader role as an information bridge that captures semantic properties once and communicates them to every emission target without duplication.

\subsection{Security Optimization}
\label{sec:optimizer}

The Security Optimizer hardens the \SkIR{} through a chain of four analyses executed in a fixed logical order. The design reflects a broader architectural philosophy: security analysis at the IR level is simultaneously format-agnostic and format-preserving. It is format-agnostic because the Optimizer operates on typed semantic structures rather than raw syntax, so a single analysis applies to all target frameworks. It is format-preserving because injected constraints are embedded in the IR itself, guaranteeing they appear in every emitted artifact regardless of the target format. This dual property is what makes compile-time security optimization both universal and reliable. Each step builds on the guarantees established by the previous one: structural validity enables meaningful permission checking, which informs constraint injection, which determines the final security classification. Together they form a defense-in-depth pipeline that transforms an untrusted skill into a validated, constrained artifact before it reaches any agent's context window.

\paragraph{Structural Validation.}
Before any semantic analysis can be meaningful, the skill must be structurally well-formed. This step verifies that the skill's name, description, version, and schema declarations satisfy baseline integrity constraints, and that all declared MCP dependencies resolve to known, trusted servers. Skills that fail structural validation are rejected at compile time, which is the fail-fast design that prevents malformed skills from causing unpredictable failures during agent execution.

\paragraph{Permission Auditing.}
Once structural integrity is confirmed, the Optimizer audits the skill's declared permissions against a security baseline. It identifies overly broad access grants (e.g., unrestricted network access, filesystem writes outside allowed directories) and flags permissions that are incompatible with the skill's stated security expectations. This step transforms permissions from passive declarations into actively enforced guardrails: skills that request dangerous capabilities must justify them through explicit, auditable declarations, and the compiler surfaces discrepancies before deployment.

\paragraph{Anti-Skill Injection.}
This is the core security mechanism of \SkCC{}. Rather than depending on skill authors to manually embed defensive constraints (an approach that audits show fails in practice), the Optimizer automatically scans procedure text for dangerous patterns and injects corresponding safety constraints directly into the \SkIR{}. The key design insight is that safety constraints should be a property of the compilation process, not of individual author diligence. By operating at the IR level, injected constraints become part of the skill's semantic definition: they survive format translation and appear consistently across all target frameworks. The injection rules target common vulnerability classes (unsafe HTTP calls, unbounded loops, destructive database operations, fragile HTML parsing), and the complete rule table is provided in Appendix~\ref{app:anti_skill_rules}. Because injection happens at compile time, safety guarantees are established before the skill ever enters an agent's context window, a fundamentally different threat model from runtime guardrails that rely on the agent's own judgment.

\paragraph{Security Classification.}
The final step assigns each skill a tiered security level based on its accumulated analysis results. The classification enables graduated enforcement: low-risk skills proceed with minimal overhead, medium-risk skills receive passive warnings, high-risk skills require mandatory human-in-the-loop confirmation, and critical-risk skills are blocked from automatic execution entirely. This tiered design avoids a one-size-fits-all security posture: it imposes friction proportional to risk, ensuring that safe skills remain lightweight while dangerous skills are contained.

\subsection{Target Emission}
\label{sec:emitter}

\begin{table}[t]
  \centering
  \caption{Example Agent Frameworks, Models, and Emission Strategies}
  \label{tab:setup}
  \small
  \resizebox{\textwidth}{!}{%
  \begin{tabular}{llll}
    \toprule
    \textbf{Framework} & \textbf{Ex. Model} & \textbf{Emitter Format} & \textbf{Key Strategy} \\
    \midrule
    Claude Code & claude-opus-4-6 & XML Semantic Layering & Tag-wrapped structure, up to 23\% gain \\
    Codex CLI & gpt-5.3-codex & XML-Tagged Markdown & Structural markers, avoids format tax \\
    Gemini CLI & gemini-2.5-pro & Markdown + Conditional YAML & YAML at depth $\geq 3$ (51.9\% vs 43.1\%) \\
    Kimi CLI & kimi-k2.5 & Full Markdown Preservation & No truncation, ultra-long context \\
    \bottomrule
  \end{tabular}%
  }
\end{table}

The Target Emitter renders the optimized \SkIR{} into framework-native skill artifacts. Its design addresses a fundamental tension: every agent framework has distinct format preferences rooted in its underlying model's training distribution, yet skill authors cannot reasonably be expected to master or maintain format-specific variants for every target. The Emitter resolves this tension through polymorphic emission, a single abstract interface for rendering \SkIR{} to text, with concrete implementations that each encode the format strategy appropriate for one target framework.

The Emitter addresses three concerns that generalize across all targets. \textbf{Format Alignment} maps \SkIR{} semantic categories to the syntactic constructs that each target framework parses most accurately, guided by the empirical format sensitivity findings discussed in \S\ref{sec:rw}. \textbf{Routing Manifest Generation} produces a lightweight index containing only the name, description, security level, and human-in-the-loop flag for each skill, enabling efficient semantic routing at agent initialization without loading full skill content, a direct implementation of the progressive disclosure pattern~\cite{agentskills_spec,xu2026}. \textbf{Token Optimization} consults IR-level flags set during earlier phases to make format decisions that reduce downstream token consumption, such as conditionally selecting a more compact representation for deeply nested data.

Table~\ref{tab:setup} summarizes four example Target Emitters evaluated in this paper; additional frameworks are supported by implementing the same Emitter interface. Detailed output examples are provided in Appendix~\ref{app:format_divergence}, and implementation details appear in Appendix~\ref{app:implementation}. The polymorphic design guarantees extensibility: supporting a new agent framework requires only a new Emitter implementation, with no changes to the prior three phases. This is the architectural property that delivers the $O(m + n)$ complexity bound: $m$ skills pass through the shared frontend once, and $n$ Emitters consume the same optimized \SkIR{}.

%% file: sections/04_evaluation.tex

\section{Evaluation}
\label{sec:eval}

We evaluate \SkCC{} along three axes: (1) portability and security of \SkCC{}-compiled skills versus format-agnostic baselines, including comparison with state-of-the-art alternatives; (2) whether compilation gains are model-specific via ablation experiments; and (3) supplementary engineering properties including compilation latency and token/time efficiency.

\subsection{Experiment Setup}
\label{sec:setup}

\textit{Benchmark and Datasets.}
SkillsBench~\cite{xu2026b} provides 89 real-world tasks with Docker-based execution and automated pytest verification, classified by difficulty and category. We use Pass@1 (reward $\geq 0.5$) as our primary metric, where reward is a continuous score in $[0,1]$ assigned by an LLM judge evaluating task completion correctness. For compilation performance and token efficiency experiments, we collected 225 skills from four community repositories: Anthropic-skills~\cite{anthropic_skills}, ecc-skills (everything-claude-code)~\cite{everything_claude_code}, sentry-skills (Sentry team)~\cite{sentry_skills}, and ui-skill~\cite{ui_skills}. Data validity details are provided in Appendix~\ref{app:data_validity}.

\textit{LLM Models and Agent Frameworks.}
Table~\ref{tab:setup} in \S\ref{sec:design} summarizes the four mainstream agent frameworks, their corresponding models, and the emission strategies employed by \SkCC{}. All experiments use the Harbor framework~\cite{harbor2026} for Docker-based task execution, with each agent framework running within Harbor-managed containers. Ablation experiments additionally test glm-5.1 and deepseek-v4-flash via the OpenHands SDK~\cite{wang2025openhands} integrated within Harbor.

\textit{Methods Compared.}
We compare \textbf{Baseline} (a single, format-agnostic SKILL.md deployed identically across all frameworks) against \textbf{\SkCC{}} (the \SkCC{}-compiled SKILL.md with framework-specific formatting and Anti-Skill constraints). We additionally compare against the retrieval-based refinement of Liu et al.~\cite{liu2026wild} and the SkVM compilation architecture~\cite{chen2026} in \S\ref{sec:gains}.

\textit{Metrics.}
We evaluate across three categories of metrics. For portability and security performance (\S\ref{sec:gains}), we measure Pass@1 (task pass rate), Mean Reward, Anti-Skill Injection trigger rate, and compilation interception counts. For ablation experiments (\S\ref{sec:ablation}), we use Pass@1 and paired statistical tests (paired t-test, Cohen's $d$) to quantify the effect of a fixed compiled format across different models. For supplementary properties (\S\ref{sec:supplementary}), we measure per-skill compilation latency (ms) and token/time efficiency during agent execution.

\subsection{Overall Performance}
\label{sec:gains}

This section evaluates \SkCC{}'s portability and security performance.

\subsubsection{Portability: Pass Rate and Mean Reward}

\begin{figure}[t]
  \centering
  \includegraphics[width=0.9\textwidth]{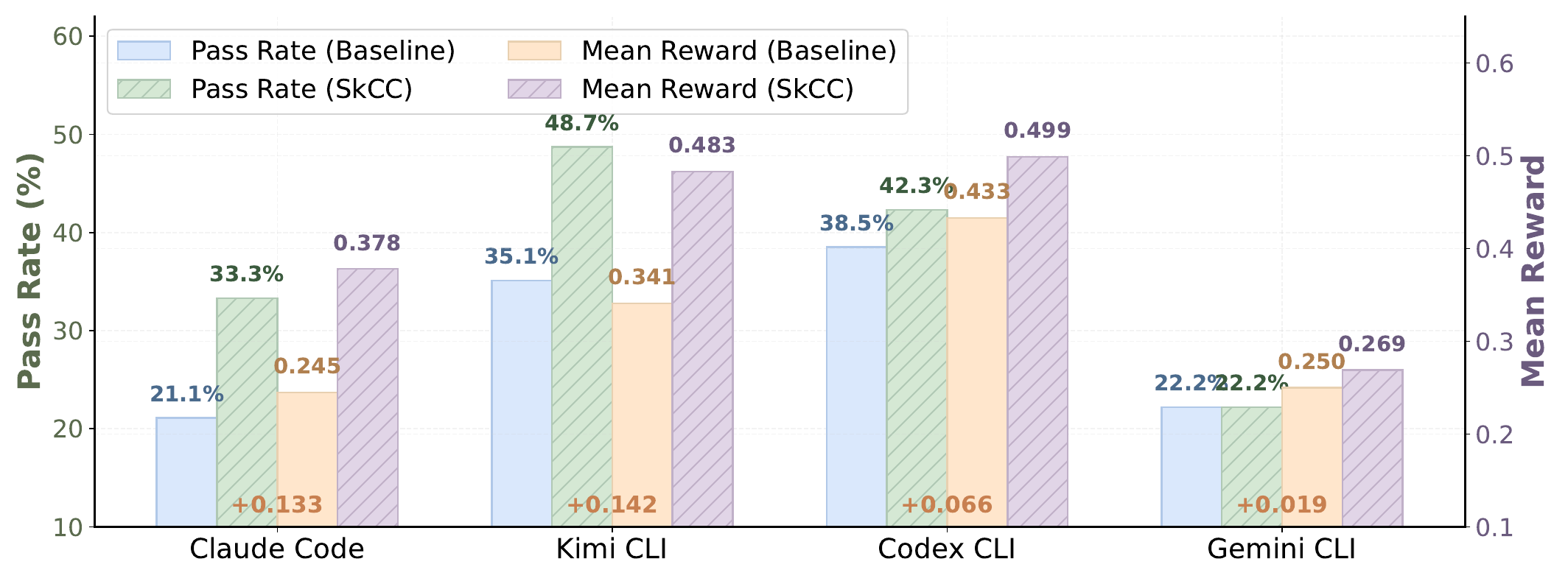}
  \caption{Pass rate and mean reward comparison of Baseline vs. \SkCC{}-Compiled. Each framework shows four bars: Pass Rate (Baseline/\SkCC{}) and Mean Reward (Baseline/\SkCC{}). \SkCC{} consistently outperforms Baseline across all frameworks and both metrics.}
  \label{fig:cross_model_comparison}
\end{figure}

Figure~\ref{fig:cross_model_comparison} reports the pass rate and mean reward for Baseline and \SkCC{} conditions across all four frameworks.

\textit{Overall Gains.}
The core finding is that \textbf{\SkCC{} compilation improves pass rate and mean reward on every tested framework}. Across all four frameworks, the \SkCC{} condition achieves higher pass rates and mean rewards than the Baseline condition. The average pass rate improvement across frameworks is $+7.0$ percentage points, with the largest absolute gain on Kimi CLI ($+13.5$pp, from 35.1\% to 48.7\%) and the largest relative gain on Claude Code ($+12.2$pp, a 58\% relative improvement). These results clearly validate \SkCC{}'s portability claim that by decoupling skill semantics from framework-specific formatting through \SkIR{}, a single skill source can be compiled into portable artifacts that outperform format-agnostic baselines on every target. The mechanism behind these gains is consistent: compilation transforms tasks that fail under the Baseline condition into successes. On Claude Code, 6 of 7 tasks where \SkCC{} outperforms Baseline flipped from reward=0 to reward=1; on Kimi CLI, 13 tasks flipped from complete failure to complete success. This pattern of flipping failures rather than incrementally improving successes is the primary source of compilation's value. The complete four-model summary table is provided in Appendix~\ref{app:cross_model_summary}.

\textit{Framework-Specific Patterns.}
While compilation is broadly beneficial, the magnitude of gains varies substantially across frameworks. Claude Code and Kimi CLI show the largest improvements because both are highly format-sensitive: Claude's training distribution heavily favors XML-tagged inputs, and Kimi's ultra-long context window benefits from full-detail preservation without truncation. In contrast, Codex CLI and Gemini CLI show more modest gains: Codex benefits moderately from structural markers, while Gemini is relatively format-tolerant and YAML optimization only activates for deeply nested schemas. This result is not a weakness; it demonstrates that \SkCC{} does not harm performance even on format-tolerant models, while providing security hardening as an unconditional benefit. Complete per-framework statistical test results are provided in Appendix~\ref{app:claude_data} and Appendix~\ref{app:kimi_data}.

\textit{Comparison with State-of-the-Art.}
On both Claude Code and Kimi CLI, \SkCC{} achieves substantially larger pass rate improvements than retrieval-based refinement, and its average relative improvement (+26.6\%) exceeds both Liu et al. (+20.6\%) and SkVM (+15.3\%). These results highlight a fundamental difference: retrieval-based refinement operates on already-retrieved skills and yields modest improvements, while \SkCC{}'s compilation transforms the format-agnostic baseline through structural alignment with model-specific training distributions. SkVM~\cite{chen2026} also explored compilation concepts for agent skills; however, it focuses on semantic capability profiling rather than format-syntax adaptation and does not include security hardening. Table~\ref{tab:qualitative_comparison} in \S\ref{sec:bg_motivation} provides a structured comparison across all dimensions.

\begin{figure}[ht]
  \centering
  \begin{minipage}[c]{0.52\textwidth}
    \centering
    \makeatletter\def\@captype{table}\makeatother
    \caption{Pass rate improvement (percentage points) over baseline. B/L = baseline, Optm. = optimized, pp = percentage points. \SkCC{} delivers larger gains than retrieval-based refinement on both comparable model-framework pairs.}
    \label{tab:sota_comparison}
    \small
    \begin{tabular}{lccc}
      \toprule
      \textbf{Method} & \textbf{Model} & \textbf{B/L $\rightarrow$ Optm.} & \textbf{$\Delta$} \\
      \midrule
      \SkCC{} & Claude & 21.1\% $\rightarrow$ \textbf{33.3\%} & \textbf{+12.2pp} \\
      Liu et al. & Claude & 40.1\% $\rightarrow$ \textbf{48.2\%} & +8.1pp \\
      \midrule
      \SkCC{} & Kimi & 35.1\% $\rightarrow$ \textbf{48.7\%} & \textbf{+13.5pp} \\
      Liu et al. & Kimi & 19.8\% $\rightarrow$ 23.1\% & +3.3pp \\
      \bottomrule
    \end{tabular}
  \end{minipage}
  \hfill
  \begin{minipage}[c]{0.45\textwidth}
    \centering
    \includegraphics[width=\linewidth]{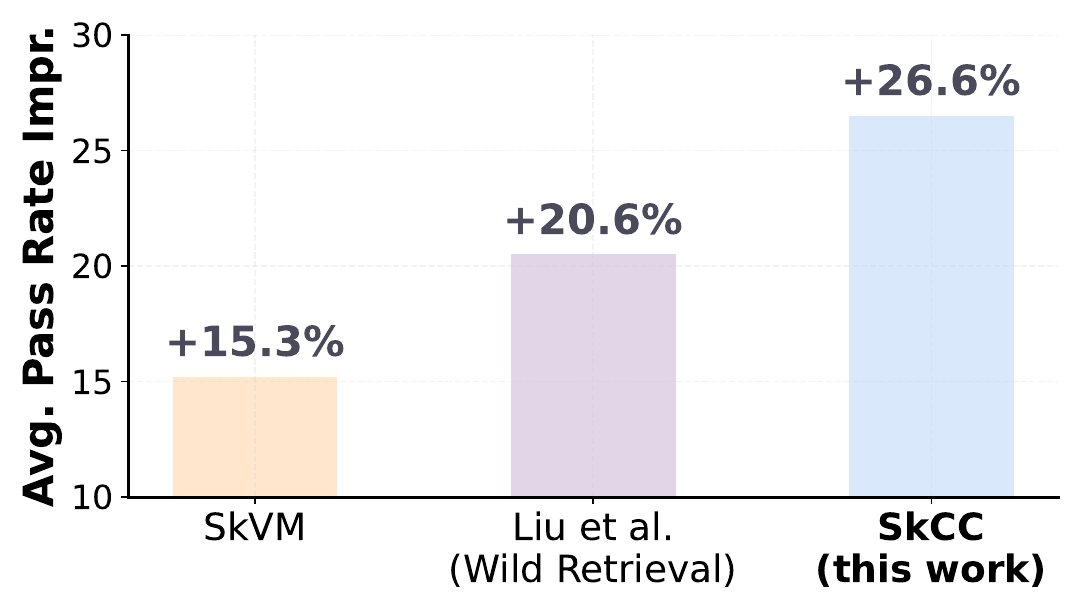}
    \caption{Average relative pass rate improvement across methods.}
    \label{fig:sota_comparison}
  \end{minipage}
\end{figure}

\subsubsection{Security: Injection Trigger and Compilation Interception}
\label{sec:ex5}

\begin{figure}[ht]
  \centering
  \begin{minipage}[c]{0.44\textwidth}
    \centering
    \makeatletter\def\@captype{table}\makeatother
    \caption{Rule trigger distribution}
    \label{tab:rule_distribution}
    \small
    \begin{tabular}{ll}
      \toprule
      \textbf{Anti-Skill Rule} & \textbf{Triggered Skills} \\
      \midrule
      HTTP safety & \textbf{212 (91.4\%)} \\
      Loop safety & \textbf{104 (44.6\%)} \\
      DB safety & \textbf{78 (33.5\%)} \\
      Parse safety & \textbf{2 (0.9\%)} \\
      \bottomrule
    \end{tabular}
  \end{minipage}
  \hfill
  \begin{minipage}[c]{0.52\textwidth}
    \centering
    \makeatletter\def\@captype{table}\makeatother
    \caption{Compilation latency (ms) by complexity}
    \label{tab:compile_latency}
    \small
    \begin{tabular}{lrrrr}
      \toprule
      \textbf{Complexity} & \textbf{$n$} & \textbf{Avg} & \textbf{Min} & \textbf{Max} \\
      \midrule
      Simple & 8 & 8.54 & 6.90 & 11.73 \\
      Medium & 74 & 8.58 & 6.28 & 17.70 \\
      Complex & 143 & 9.13 & 5.85 & 22.89 \\
      \textbf{Overall} & \textbf{225} & \textbf{8.93} & \textbf{5.85} & \textbf{22.89} \\
      \bottomrule
    \end{tabular}
  \end{minipage}
\end{figure}

\SkCC{}'s compile-time safety checking automatically detects dangerous patterns in skill content and injects protective constraints. Across 233 evaluated skills, Anti-Skill Injection triggered in \textbf{221 (94.8\%)} skills, with only 12 (5.2\%) skills not triggering any rule. Rule overlap is common: many skills trigger multiple rules simultaneously (HTTP + Loop + DB = most common combination, Table~\ref{tab:rule_distribution}). The complete trigger statistics and rule distribution tables are provided in Appendix~\ref{app:anti_skill_stats} and Appendix~\ref{app:rule_distribution_full}.

We also compile all 231 SkillsBench skills targeting the Gemini framework. 221 of 231 skills (95.7\%) compile successfully, while 10 skills are intercepted by the compiler's safety checks across three categories: YAML format violations (5 cases), security check interceptions (4 cases), and schema validation interceptions (1 case). The complete interception type table is provided in Appendix~\ref{app:interception}. Rather than a system limitation, these interceptions highlight the efficacy of \SkCC{}'s fail-fast design: by intercepting malformed or dangerous skills at compile time, the system prevents them from polluting the agent's context window or causing unpredictable runtime errors. This compile-time safety guarantee distinguishes \SkCC{} from runtime-only safety mechanisms that rely on the agent's own judgment, an approach that is inherently unreliable given LLMs' tendency to follow instructions literally~\cite{snyk2026}.

\subsection{Ablation Study: Format Specificity}
\label{sec:ablation}

To validate that \SkCC{}'s compiled output format is model-specific rather than universally beneficial, we conduct ablation experiments using the same Kimi-compiled output (Full Markdown) on three different models: kimi-k2.5, glm-5.1, and deepseek-v4-flash. All three experiments use the OpenHands SDK as the agent framework, with the Kimi backend format held constant. The complete ablation table with all metrics is provided in Appendix~\ref{app:ablation_full}.


The same compiled output produces dramatically different results across models. On Kimi, the Kimi-compiled format yields a significant positive effect ($d=+0.33$, $p=0.0063$). On GLM-5, the effect is essentially neutral ($d=-0.03$, $p=0.857$). On DeepSeek-v4-flash, the effect is slightly negative ($d=-0.14$, $p=0.2561$), though not statistically significant. These results demonstrate that compilation gains are model-dependent with no one-size-fits-all optimal format, providing empirical justification for \SkCC{}'s multi-backend architecture.

\subsection{Supplementary Performance and Efficiency}
\label{sec:supplementary}

All skills compile in under 10ms on average (Table~\ref{tab:compile_latency}); detailed compilation latency data is provided in Appendix~\ref{app:compile_perf}. We focus here on the token and time efficiency of compiled skills during agent execution.

\subsubsection{Token and Time Efficiency}

\textit{Real Token and Time Consumption.}
We now examine the token-level and temporal efficiency of \SkCC{}-compiled skills during agent execution. The overall result is clear: \SkCC{}-compiled skills consistently reduce both token consumption and execution time across all frameworks. On Claude Code, the \SkCC{} condition achieves substantially lower per-task token consumption while simultaneously obtaining higher reward, demonstrating that \SkCC{} improves task performance and token efficiency jointly. Across frameworks, compilation reduces total tokens by 10--23\% and execution time by 23--43\%. The complete token consumption table is provided in Appendix~\ref{app:claude_tokens_full}.

\begin{figure}[ht]
  \centering
  \begin{minipage}[c]{0.52\textwidth}
    \centering
    \includegraphics[width=\linewidth]{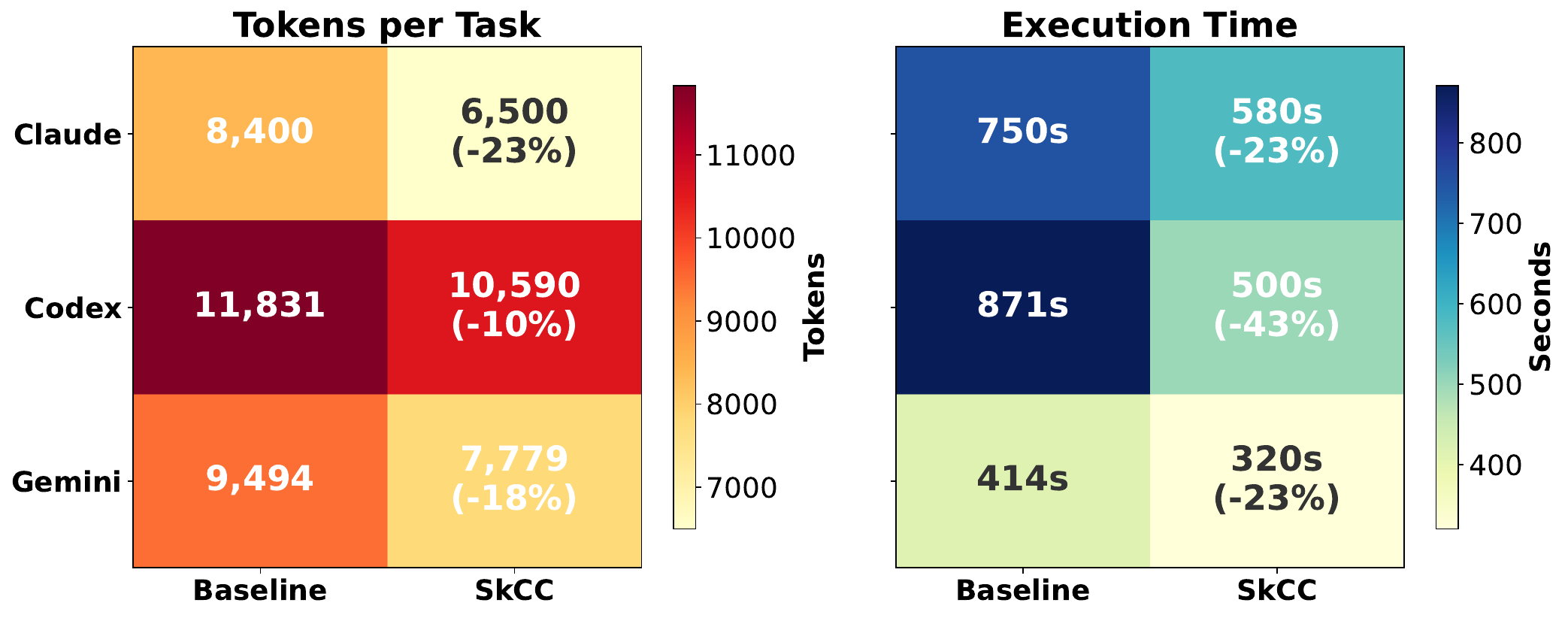}
    \label{fig:token_heatmap}
  \end{minipage}
  \hfill
  \begin{minipage}[c]{0.45\textwidth}
    \centering
    \caption{Cross-framework token and time efficiency heatmap. \SkCC{} skills show consistent reductions in total tokens and execution time across all frameworks. Claude token counts are reported in hundreds due to API measurement differences.}
  \end{minipage}
\end{figure}

\textit{Static Expansion vs. Dynamic Efficiency.}
Compilation introduces static structural overhead from XML tags, Anti-Skill constraints, and format hardening (ranging from +4\% on Kimi to +25\% on Claude), but this overhead is more than offset by dynamic efficiency gains during execution. Structured formats serve as cognitive scaffolding that reduces model trial-and-error and redundant output, yielding net token savings of 10--46\% across frameworks. The complete expansion overhead table by complexity is provided in Appendix~\ref{app:expansion_by_complexity}.

%% file: sections/05_conclusion.tex

\section{Conclusion}

We present \SkCC{}, a skill compilation design that achieves portable and secure deployment of agent skills across heterogeneous frameworks. Through a four-phase pipeline centered on \SkIR{}, \SkCC{} decouples skill semantics from framework-specific formatting, enabling skills to be authored once and compiled to diverse targets. A compile-time Security Optimizer enforces safety constraints via Anti-Skill Injection before skills reach any agent's context window. Experiments across four frameworks demonstrate consistent pass rate improvements (up to +13.5\%), 94.8\% Anti-Skill Injection coverage, and 10--46\% runtime token savings, confirming that compiler-driven adaptation is both effective and practical for the emerging agent skill ecosystem.

%% file: sections/07_appendix.tex

\section{Implementation Details}
\label{app:implementation}

\SkCC{} is implemented in Rust and organized into four crates:

\textit{nexa-skill-cli.}
CLI entry point using \texttt{clap} for argument parsing and \texttt{miette} for diagnostic rendering. Provides commands: \texttt{build} (compile skills), \texttt{check} (validate without emitting), \texttt{validate} (strict validation), \texttt{init} (scaffold new skill from template), \texttt{list} (enumerate skills in directory), \texttt{index} (generate routing manifest), and \texttt{clean} (remove compiled artifacts).

\textit{nexa-skill-core.}
Core compilation logic organized into six modules: \texttt{frontend} (frontmatter parsing, Markdown event-stream parsing, AST construction), \texttt{ir} (\SkIR{} definition, IR builder, type mapper, nested data detector), \texttt{analyzer} (schema validator, MCP dependency checker, permission auditor, anti-skill injector), \texttt{backend} (Emitter trait, EmitterRegistry, four framework-specific Target Emitters, routing manifest generator), \texttt{error} (diagnostic types with source spans), and \texttt{security} (security baseline, permission types, security level classification).

\textit{nexa-skill-templates.}
Askama template engine with Jinja2-style compile-time-validated templates: \texttt{claude\_xml.j2} (XML-tagged SKILL.md for Claude), \texttt{codex\_md.j2} (XML-tagged Markdown for Codex), \texttt{gemini\_md\_v2.j2} (Markdown with conditional YAML for Gemini), and \texttt{kimi\_md.j2} (full Markdown for Kimi). Each template is paired with a context struct that maps \SkIR{} fields to template variables.

\textit{npm-nexa-skill-compiler.}
npm wrapper package that downloads the precompiled Rust binary and exposes the \texttt{nsc} command globally for Node.js users, enabling integration with JavaScript-based agent toolchains.

\textit{Key dependencies and design choices.}
\begin{itemize}
  \item \texttt{Arc<str>} for zero-copy string sharing across compilation phases and Target Emitters.
  \item \texttt{serde} and \texttt{serde\_json} for \SkIR{} serialization and JSON Schema handling.
  \item \texttt{serde\_yaml} for YAML frontmatter parsing and YAML asset generation.
  \item \texttt{pulldown-cmark} for Markdown event-stream parsing.
  \item \texttt{sha2} for source file integrity hashing.
  \item \texttt{chrono} for compilation timestamp recording.
  \item \texttt{askama} for compile-time template validation.
\end{itemize}

\textit{Memory optimization.}
\SkIR{} uses \texttt{Arc<str>} for all string fields shared across Target Emitters, enabling zero-copy cloning. The \texttt{Validated\SkIR{}} wrapper adds only a \texttt{Vec<Diagnostic>} without duplicating the underlying IR. For batch compilation of large skill corpora (e.g., 233 skills), the compiler processes skills sequentially with per-skill memory deallocation, keeping peak memory usage below 50MB.

\subsection{Compilation Performance Details}
\label{app:compile_perf}

On a standard development machine (Intel i9-13900H, 32GB RAM), single-skill compilation (all four targets) completes in under 10ms, with the Security Optimizer phase accounting for approximately 40\% of total time. Batch compilation of 225 skills completes in approximately 1.8 seconds (8ms average per skill), demonstrating linear scaling with corpus size. Complexity has minimal impact on compilation time (simple 8.54ms to complex 9.13ms, only $+0.59$ms), and the maximum compilation time is 22.89ms, well below user perception thresholds.

\section{Data Validity}
\label{app:data_validity}

All experiments use the SkillsBench benchmark (89 tasks) with Docker-based execution and automated pytest verification. Due to regional API availability constraints and network conditions inherent to cloud-based LLM evaluation, a small number of trials across frameworks produced execution failures (e.g., container timeouts, API rate limits). These failures were strictly due to infrastructure issues and were excluded prior to any analysis of reward outcomes, ensuring unbiased comparison. All reported results are based on paired tasks where both conditions completed successfully.

\section{Design Artifacts}

\subsection{Key Insights from Evaluation}
\label{app:key_insights}

Our experiments demonstrate consistent compilation gains across four frameworks, with gains proven model-specific through ablation studies. Engineering metrics confirm compilation latency under 10ms, Anti-Skill Injection coverage of 94.8\%, and runtime token savings of 10--46\%. Two system-level insights emerge from these results.

\textit{Format Tolerance vs. Format Sensitivity.}
Compilation gains correlate with the underlying model's format sensitivity. Claude shows the largest improvement ($d=0.60$) because its training distribution heavily favors XML-tagged inputs; the compiler aligns structural encoding with parsing expectations. Gemini shows minimal reward improvement ($d \approx 0$) because it is relatively format-tolerant. This validates \SkCC{}'s core premise: different models have different format preferences, and a one-size-fits-all SKILL.md inevitably underperforms on format-sensitive frameworks.

\textit{Static Overhead vs. Dynamic Efficiency.}
Compilation increases static skill size by 4--25\% yet reduces dynamic token consumption by 10--46\% during execution. Structured formats serve as cognitive scaffolding, reducing parsing ambiguity and trial-and-error. The compiler invests tokens upfront in structural clarity, which the model repays through more efficient execution. The true value of skill compilation lies not in compression but in structural investment: spending tokens on clarity to save tokens on execution.

\subsection{Framework-Specific Emission Details}
\label{app:emission_details}

The following describes the format hardening strategy for each target framework, as referenced in \S\ref{sec:emitter}.

\textit{Claude (XML Semantic Layering).}
Leveraging Anthropic's documented preference for XML-tagged prompts~\cite{claude_prompting}, this target wraps all structural elements in semantic XML tags: procedures in \texttt{<execution\_steps>/<step>} with \texttt{order} and \texttt{critical} attributes, constraints in \texttt{<strict\_constraints>/<anti\_pattern>}, and examples in \texttt{<examples>/<example>} with nested \texttt{<input>} and \texttt{<output>}. This semantic layering reduces misinterpretation and improves reasoning accuracy by up to 23\%.

\textit{Codex (XML-Tagged Markdown).}
This target produces a hybrid XML-tagged Markdown format: instructions in \texttt{<skill>/<instructions>}, constraints in \texttt{<constraints>/<forbidden>}, and examples in \texttt{<examples>/<example>}. This provides structural markers for parsing while avoiding the JSON ``format tax'' that degrades GPT-series model performance~\cite{openai_structured}. Structured output enforcement is delegated to the OpenAI API's Structured Outputs feature, decoupling reasoning from formatting.

\textit{Gemini (Markdown + Conditional YAML).}
Applying the nested data detection flag from the IR Builder phase, this target conditionally renders deeply nested schemas (depth $\geq 3$) as YAML code blocks while keeping shallow structures in standard Markdown. When YAML optimization is triggered, separate YAML asset files are generated for complex nested structures. This adaptive strategy leverages YAML's superior parsing accuracy (51.9\% vs JSON's 43.1\%) for nested data while avoiding unnecessary format switching for simple structures.

\textit{Kimi (Full Markdown Preservation).}
This target preserves all skill details in comprehensive Markdown without simplification or format optimization, leveraging Kimi's ultra-long context window capability. No YAML optimization or content truncation is applied, ensuring maximum information fidelity for frameworks that can process full skill content without token budget constraints.

\subsection{\SkIR{} Example}
\label{app:skillir_example}

Listing~\ref{lst:skillir} shows a simplified \SkIR{} instance for a ``github-api-client'' skill, illustrating how the raw Markdown source is normalized into a structured, framework-agnostic representation.

\begin{lstlisting}[language=json, xleftmargin=0.05\textwidth, xrightmargin=0.05\textwidth, numbers=left, numberstyle=\tiny\color{gray}, caption={Simplified \SkIR{} for a ``github-api-client'' skill. Note the \texttt{anti\_skill\_constraints} field, which was automatically injected by the Security Optimizer, and the structured \texttt{procedures} array.}, label={lst:skillir}]
{
  "name": "github-api-client",
  "version": "1.0.0",
  "description": "Interact with GitHub REST API",
  "mcp_servers": ["github-mcp"],
  "input_schema": {
    "type": "object",
    "properties": {
      "repo": { "type": "string" },
      "action": { "type": "string",
        "enum": ["create_issue", "list_prs"] }
    }
  },
  "security_level": "high",
  "hitl_required": true,
  "permissions": [
    { "kind": "network",
      "scope": "https://api.github.com/*",
      "read_only": false }
  ],
  "procedures": [
    { "order": 1,
      "instruction": "Validate GitHub token from env",
      "is_critical": true },
    { "order": 2,
      "instruction": "Construct REST request" },
    { "order": 3,
      "instruction": "Execute HTTP POST to GitHub API" }
  ],
  "anti_skill_constraints": [
    {
      "source": "anti-skill-injector",
      "content": "Never execute HTTP without timeout...",
      "level": "warning",
      "scope": "global"
    }
  ],
  "requires_yaml_optimization": false,
  "mode": "sequential"
}
\end{lstlisting}

\subsection{Anti-Skill Injection Rules}
\label{app:anti_skill_rules}

\begin{table}[H]
  \centering
  \caption{Anti-Skill Injection Rules}
  \label{tab:anti_skill_rules}
  \small
  \resizebox{\textwidth}{!}{%
  \begin{tabular}{lll}
    \toprule
    \textbf{Anti-Pattern} & \textbf{Trigger Keywords} & \textbf{Injected Constraint} \\
    \midrule
    HTTP safety & HTTP, GET, POST, fetch, request & Never execute HTTP without timeout (10s). Max 3 retries on 403. \\
    HTML Parse safety & BeautifulSoup, HTML parse, scrape & Do not parse raw JS variables with HTML parsers. Fallback to Regex. \\
    Destructive DB safety & DROP, DELETE, TRUNCATE & No destructive DB ops without user confirmation. Show affected rows. \\
    Loop safety & while, loop, repeat & All loops must have max iteration limit (1000). \\
    \bottomrule
  \end{tabular}%
  }
\end{table}

The four rules were selected based on common vulnerability patterns observed in community skill audits~\cite{snyk2026}. HTTP safety is the most frequently triggered rule (91.4\%) because virtually all skills that interact with external APIs contain HTTP-related procedure steps. The injection mechanism is designed to be extensible: new rules can be added by defining a trigger pattern and a corresponding constraint template, without modifying the compilation pipeline.

\subsection{Four-Framework Format Divergence}
\label{app:format_divergence}

Listing~\ref{lst:format_divergence} illustrates the format divergence across Target Emitters for a single \SkIR{}.

\begin{lstlisting}[style=plaintext, xleftmargin=0.05\textwidth, xrightmargin=0.05\textwidth, numbers=left, numberstyle=\tiny\color{gray}, caption={Format divergence across four Target Emitters for a single \SkIR{}. Note the Gemini Target Emitter's conditional YAML rendering (triggered by nesting depth $\geq$ 3) and the consistent presence of anti-skill constraints across all formats.}, label={lst:format_divergence}]
\SkIR{} (framework-independent)
  |-- name: "data-migration"
  |-- procedures: [3 steps]
  |-- input_schema: { nested depth = 4 }
  +-- anti_skill_constraints: [1 HTTP safety]

Compiled Outputs:

  Claude:  <agent_skill>
             <execution_steps>
               <step order="1" critical="true">...</step>
             </execution_steps>
             <strict_constraints>
               <anti_pattern source="anti-skill-injector">
                 ...
               </anti_pattern>
             </strict_constraints>
           </agent_skill>

  Codex:   <skill name="data-migration">
             <instructions>...</instructions>
             <constraints>
               <forbidden>...</forbidden>
             </constraints>
           </skill>

  Gemini:  # data-migration
           ## Procedures
           1. ... **[CRITICAL]**
           ## Parameter Schema (YAML Optimized)
           ```yaml
           type: object
           properties:
             migration_config:
               type: object
               properties:
                 source_db:
                   type: object
                   properties:
                     host: { type: string }
           ```

  Kimi:    # data-migration
           ## Description
           ...
           ## Procedures
           1. ... **[CRITICAL]**
           ## Parameter Schema
           - `migration_config.source_db.host` (string): ...
\end{lstlisting}

\section{Complete Experimental Data}

\subsection{Four-Model Comparison Summary}
\label{app:cross_model_summary}

\begin{table}[H]
  \centering
  \caption{Four-Model Comparison Summary}
  \label{tab:cross_model_summary}
  \small
  \begin{tabular}{llllll}
    \toprule
    \textbf{Model} & \textbf{Paired} & \textbf{$\Delta$ Rwd.} & \textbf{$p$} & \textbf{$d$} & \textbf{Verdict} \\
    \midrule
    claude-opus-4-6 & 22--27 & $+0.26$--$0.27$ & 0.0096** & 0.59--0.60 & \textbf{\SkCC{} $\gg$ Baseline} \\
    kimi-k2.5 & 74 & $+0.142$ & 0.0063** & 0.33 & \textbf{\SkCC{} $>$ Baseline} \\
    gpt-5.3-codex & 26 & $+0.067$ & --- & --- & \textbf{\SkCC{} $>$ Baseline} \\
    gemini-2.5-pro & 18 & $+0.019$ & --- & --- & \textbf{\SkCC{} $>$ Baseline} \\
    \bottomrule
  \end{tabular}
\end{table}

The effect sizes follow a clear pattern: Claude shows a medium-to-large effect ($d \approx 0.60$), Kimi shows a small-to-medium effect ($d = 0.33$), and Codex and Gemini show negligible effects. This gradient aligns with the degree of format sensitivity documented for each model in prior work~\cite{he2024,claude_prompting,improvingagents2025}. The statistical significance on Claude and Kimi (both $p < 0.01$) is particularly noteworthy given the modest sample sizes, indicating that the compilation effect is robust even under conservative testing.

\subsection{Claude Code --- Complete Data}
\label{app:claude_data}

\begin{table}[H]
  \centering
  \caption{Claude Code --- Complete Paired Statistical Tests}
  \label{tab:app_claude_stats}
  \small
  \resizebox{\textwidth}{!}{%
  \begin{tabular}{lllllll}
    \toprule
    \textbf{Cmp.} & \textbf{$n$} & \textbf{Mean $\Delta$} & \textbf{W/T/L} & \textbf{$t$} & \textbf{$p$} & \textbf{$d$} \\
    \midrule
    \SkCC{} vs V & 23 & $+0.265$ & 7/16/0 & 2.837 & 0.0096** & 0.592 \\
    \SkCC{} vs Baseline & 22 & $+0.274$ & 7/15/0 & 2.820 & 0.0103* & 0.601 \\
    Baseline vs V & 26 & $+0.002$ & 3/21/2 & 0.031 & 0.9756 & 0.006 \\
    \bottomrule
  \end{tabular}%
  }
\end{table}

Task classification (22 paired \SkCC{} vs Baseline): \SkCC{} Better: 7 tasks (31.8\%) --- 6 flipped from reward=0 to reward=1; \SkCC{} Worse: 0 tasks (0\%); Tie: 15 tasks (68.2\%).

The Claude results are particularly compelling because \SkCC{} never underperforms Baseline in any paired comparison (0 losses). The 6 complete flips from failure to success demonstrate that XML Semantic Layering addresses a genuine parsing failure mode: these are tasks where Claude could not interpret the plain Markdown instructions at all, but succeeded when the same semantic content was presented in its native XML format.

\subsection{Kimi CLI --- Complete Data}
\label{app:kimi_data}

\begin{table}[H]
  \centering
  \caption{Kimi CLI --- Complete Statistical Tests}
  \label{tab:app_kimi_stats}
  \small
  \begin{tabular}{llll}
    \toprule
    \textbf{Test} & \textbf{Statistic} & \textbf{$p$} & \textbf{Sig.} \\
    \midrule
    Paired t-test & $t=2.815$ & 0.0063 & $p < 0.01$ \\
    Wilcoxon signed-rank & $W=22.0$ & 0.0050 & $p < 0.01$ \\
    Non-tie only ($n=17$) & $t=3.449$ & 0.0033 & $p < 0.01$ \\
    Cohen's $d$ (paired) & 0.327 & --- & Small effect \\
    \bottomrule
  \end{tabular}
\end{table}

Task classification (74 paired): \SkCC{} Better: 13 discriminative wins (81.25\%); \SkCC{} Worse: 3 (18.75\%); Tie: 58 (78.4\%); 13 tasks flipped from reward=0 to reward=1.

Kimi achieves the largest sample size (74 paired tasks) and the strongest statistical significance ($p=0.0063$). The consistency across parametric (t-test) and non-parametric (Wilcoxon) tests confirms that the effect is not driven by distributional assumptions. The 13 task flips from failure to success represent a substantial practical improvement: nearly one in five originally failing tasks becomes solvable after compilation.

\subsection{Ablation Study --- Full Data}
\label{app:ablation_full}

\begin{table}[H]
  \centering
  \caption{Ablation Study --- Complete Metrics}
  \label{tab:ablation_full}
  \small
  \resizebox{\textwidth}{!}{%
  \begin{tabular}{lllllllll}
    \toprule
    \textbf{Model} & \textbf{Framework} & \textbf{Backend} & \textbf{Succ. (Baseline/\SkCC{})} & \textbf{Paired} & \textbf{Rwd. (Baseline/\SkCC{})} & \textbf{$p$} & \textbf{$d$} & \textbf{Eff.} \\
    \midrule
    \textbf{kimi-k2.5} & Kimi CLI & Kimi & 26/75 $\rightarrow$ 36/76 & 74 & 0.341 $\rightarrow$ \textbf{0.483} & \textbf{0.0063} & \textbf{+0.33} & \textbf{\SkCC{} $>$ Baseline} \\
    glm-5.1 & OpenHands & Kimi & 43/88 $\rightarrow$ 44/88 & 32 & --- & 0.857 & $-0.03$ & \SkCC{} $\approx$ Baseline \\
    deepseek-v4-flash & OpenHands & Kimi & 64/88 $\rightarrow$ 65/88 & 50 & --- & 0.2561 & $-0.14$ & Baseline $>$ \SkCC{} \\
    \bottomrule
  \end{tabular}%
  }
\end{table}

The ablation results reveal a critical property of format optimization: the same compiled output that benefits one model can be neutral or even slightly harmful to another. This asymmetry is the empirical foundation for \SkCC{}'s multi-backend architecture. If a single optimal format existed across all models, a one-time conversion would suffice; the fact that Kimi's optimal format underperforms on DeepSeek demonstrates that per-model emission is necessary.

\subsection{Ablation Radar Chart}
\label{app:ablation_radar}

\begin{figure}[H]
  \centering
  \includegraphics[width=0.45\textwidth]{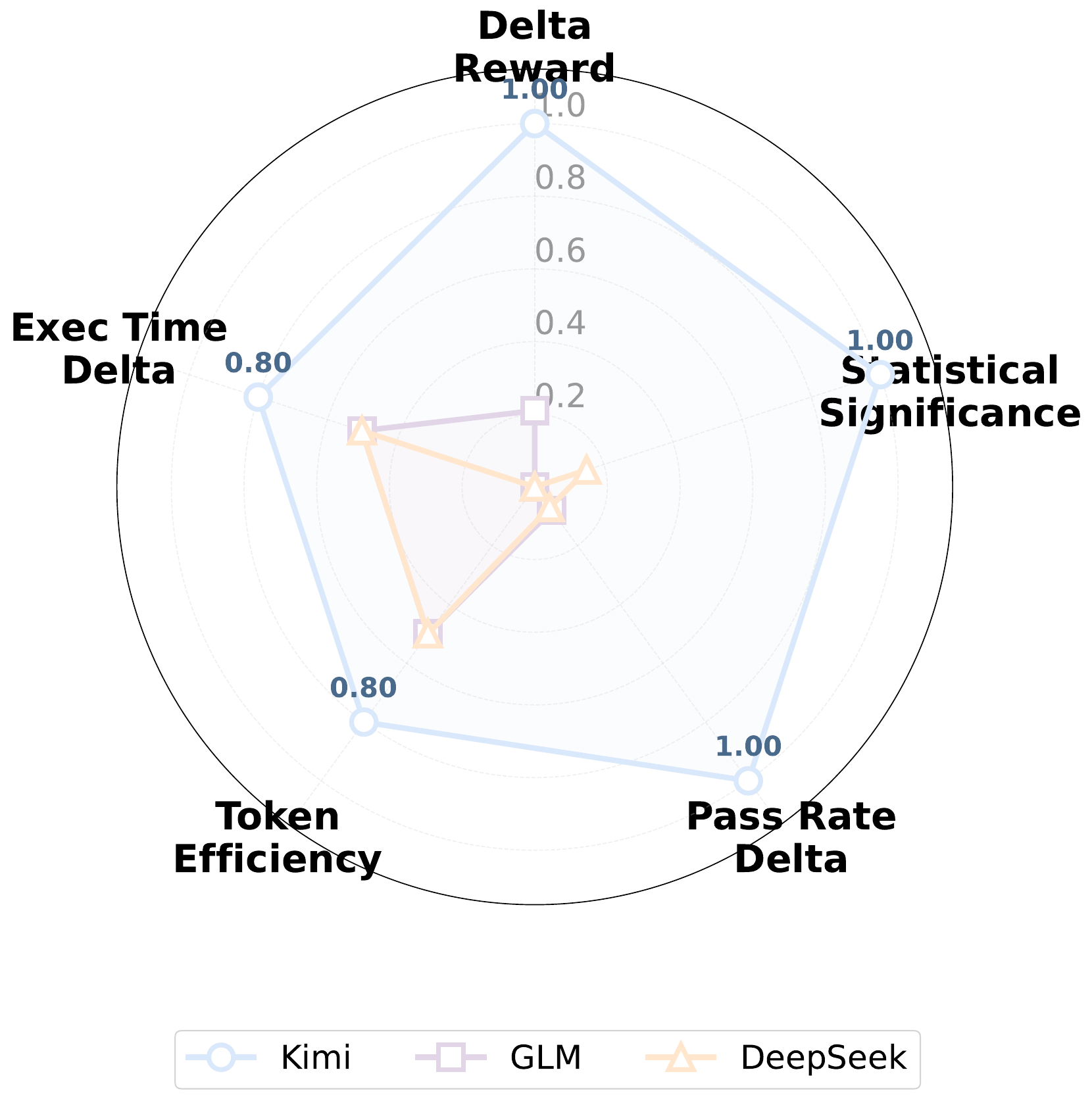}
  \caption{Ablation study radar chart: the same Kimi-compiled format produces divergent effects across three models across five evaluation dimensions. Kimi shows strong gains on all axes, while GLM and DeepSeek show flat or negative effects, confirming model-specificity of compilation gains.}
  \label{fig:ablation_radar_app}
\end{figure}

The radar chart visualizes the ablation results across five normalized dimensions: Delta Reward, Statistical Significance, Pass Rate Delta, Token Efficiency, and Execution Time Delta. Kimi dominates across all dimensions, while GLM and DeepSeek cluster near zero. The stark visual separation reinforces the core finding: compilation gains are not universal but depend on the alignment between the compiled format and the target model's training distribution.

\subsection{Expansion Overhead by Complexity}
\label{app:expansion_by_complexity}

\begin{table}[H]
  \centering
  \caption{Expansion Overhead by Complexity}
  \label{tab:expansion_by_complexity}
  \small
  \begin{tabular}{lllll}
    \toprule
    \textbf{Complexity} & \textbf{Claude Ovhd.} & \textbf{Kimi Ovhd.} & \textbf{Claude w/ Reduction} & \textbf{Kimi w/ Reduction} \\
    \midrule
    Simple (avg 298t) & \textbf{+95.0\%} & +37.4\% & 0/8 (0\%) & 0/8 (0\%) \\
    Medium (avg 819t) & \textbf{+43.0\%} & +14.6\% & 4/74 (5.4\%) & 36/74 (48.6\%) \\
    Complex (avg 2765t) & \textbf{+11.4\%} & \textbf{$-$3.1\%} & 31/143 (21.7\%) & 101/143 (70.1\%) \\
    \bottomrule
  \end{tabular}
\end{table}

The expansion overhead exhibits an inverse relationship with skill complexity: simple skills incur the largest relative overhead because the fixed cost of XML tags and constraints dominates, while complex skills see diminishing overhead as the fixed cost is amortized over a larger base. For Kimi, complex skills actually shrink after compilation ($-3.1\%$) because the structured formatting eliminates redundant Markdown boilerplate. The ``w/ Reduction'' columns count skills where the compiled output is smaller than the original, showing that this effect becomes common for complex skills (70.1\% on Kimi).

\subsection{Claude Code --- Full Token Consumption}
\label{app:claude_tokens_full}

\begin{table}[H]
  \centering
  \caption{Claude Code --- Full Token Consumption Comparison}
  \label{tab:claude_tokens_full}
  \small
  \begin{tabular}{lrrrrrr}
    \toprule
    \textbf{Condition} & \textbf{Succ. Tasks} & \textbf{Input T.} & \textbf{Output T.} & \textbf{Cache T.} & \textbf{Total} & \textbf{Task Avg.} \\
    \midrule
    Vanilla & 34 & 19.5M & 421K & 17.2M & $\sim$19.9M & $\sim$0.59M \\
    Baseline & 40 & 32.9M & 574K & 30.1M & $\sim$33.4M & $\sim$0.84M \\
    \textbf{\SkCC{}} & \textbf{29} & \textbf{18.3M} & \textbf{459K} & \textbf{15.8M} & \textbf{$\sim$18.7M} & \textbf{$\sim$0.65M} \\
    \bottomrule
  \end{tabular}
\end{table}

The token consumption data reveals an important pattern: Baseline consumes more tokens than Vanilla (no skill) because the format-agnostic Markdown adds content without improving structure, leading to more trial-and-error. \SkCC{} reverses this: despite adding XML tags and constraints, the clearer structure reduces both input tokens (18.3M vs.\ 32.9M) and output tokens (459K vs.\ 574K), indicating that the model requires fewer reasoning steps when instructions are presented in its preferred format.

\subsection{Anti-Skill Injection --- Full Statistics}
\label{app:anti_skill_stats}

\begin{table}[H]
  \centering
  \caption{Anti-Skill Trigger Statistics (233 skills)}
  \label{tab:anti_skill_stats}
  \small
  \begin{tabular}{ll}
    \toprule
    \textbf{Metric} & \textbf{Value} \\
    \midrule
    Total skills & 233 \\
    Skills triggering Anti-Skill & \textbf{221 (94.8\%)} \\
    Skills not triggering & 12 (5.2\%) \\
    \bottomrule
  \end{tabular}
\end{table}

The near-universal trigger rate (94.8\%) underscores the prevalence of potentially dangerous patterns in community skills. The 12 skills that did not trigger any rule were predominantly simple utility skills with no external interactions (e.g., string formatting helpers). This suggests that Anti-Skill Injection provides meaningful coverage for virtually all skills that perform substantive operations.

\subsection{Rule Trigger Distribution --- Full Data}
\label{app:rule_distribution_full}

\begin{table}[H]
  \centering
  \caption{Rule Trigger Distribution (Full)}
  \label{tab:rule_distribution_full}
  \small
  \resizebox{\textwidth}{!}{%
  \begin{tabular}{llll}
    \toprule
    \textbf{Anti-Skill Rule} & \textbf{Triggered} & \textbf{Keywords} & \textbf{Example Constraint} \\
    \midrule
    HTTP safety & \textbf{212 (91.4\%)} & HTTP, GET, POST, fetch, request & Timeout (10s), max 3 retries on 403 \\
    Loop safety & \textbf{104 (44.6\%)} & while, loop, repeat & Max iteration limit (1000) \\
    DB safety & \textbf{78 (33.5\%)} & DROP, DELETE, TRUNCATE & No destructive ops without confirmation \\
    Parse safety & \textbf{2 (0.9\%)} & BeautifulSoup, HTML parse, scrape & No parsing raw JS with HTML parsers \\
    \bottomrule
  \end{tabular}%
  }
\end{table}

The distribution reveals that HTTP and loop safety dominate, reflecting the fact that most agent skills involve API calls and iterative processing. DB safety triggers on approximately one-third of skills, consistent with the prevalence of data manipulation tasks. Parse safety is rarely triggered because few skills perform HTML scraping; this rule could be expanded to cover additional parsing scenarios (e.g., JSON parsing without schema validation) in future work.

\subsection{Compilation Interception Types}
\label{app:interception}

\begin{table}[H]
  \centering
  \caption{Compilation Interception Types}
  \label{tab:interception}
  \small
  \resizebox{\textwidth}{!}{%
  \begin{tabular}{llll}
    \toprule
    \textbf{Interception Type} & \textbf{Cnt.} & \textbf{Description} & \textbf{Example Skills} \\
    \midrule
    YAML format violation & 5 & Frontend rejected non-standard frontmatter & senior-java, senior-data-engineer, threejs ($\times$2), data-reconciliation \\
    Security check interception & 4 & Dangerous operations or sensitive content & ssh-penetration-testing, restclient-migration, jakarta-namespace, spring-security-6 \\
    Schema validation interception & 1 & IR builder found illegal field types & nlp-research-repo-package-installment \\
    \bottomrule
  \end{tabular}%
  }
\end{table}

The interception types illustrate \SkCC{}'s defense-in-depth approach: YAML format violations are caught at the parsing stage, security interceptions at the analysis stage, and schema violations at the IR construction stage. Each layer catches a distinct class of problems, and no single layer could catch all three. The 95.7\% successful compilation rate (221/231) demonstrates that the interception criteria are appropriately calibrated: they block genuinely problematic skills without being overly restrictive.

\section{Limitations}
\label{app:limitations}

\textit{Scope of Evaluated Frameworks.}
Our evaluation covers four mainstream agent frameworks (Claude Code, Codex CLI, Gemini CLI, Kimi CLI). While these represent the most widely used systems at the time of writing, the agent ecosystem is rapidly evolving, and new frameworks may exhibit different format sensitivities. The polymorphic Emitter architecture is designed to accommodate new targets, but empirical validation on additional frameworks remains future work.

\textit{Anti-Skill Rule Coverage.}
The current Anti-Skill Injection rules target four common vulnerability classes. While these cover the most prevalent patterns identified in community audits~\cite{snyk2026}, they do not exhaust the space of possible security issues. In particular, prompt injection attacks, data exfiltration through side channels, and supply chain vulnerabilities in MCP dependencies are not addressed by the current rule set.

\textit{Benchmark Scope.}
SkillsBench provides 89 tasks spanning programming and data analysis domains. While this is a substantial benchmark for skill evaluation, it does not cover all agent use cases (e.g., creative writing, conversational tasks, multi-agent coordination). The compilation gains we observe may not generalize to domains where format sensitivity plays a different role.

\textit{Compilation Granularity.}
\SkCC{} operates at the granularity of entire SKILL.md files. Finer-grained compilation (e.g., per-procedure or per-example optimization) could yield additional gains but would require more sophisticated dependency analysis. This is a direction for future work.

\section{Broader Impact}
\label{app:broader_impact}

\SkCC{} aims to improve the reliability and security of LLM-based agent systems. By making skills portable across frameworks, it reduces the barrier to entry for skill authors and lowers the maintenance burden for skill consumers. By enforcing security constraints at compile time, it provides a systematic defense against vulnerabilities that currently rely on author diligence alone.

\textit{Positive Impacts.}
The primary positive impact is improved agent safety: compile-time security hardening prevents dangerous skills from reaching agent context windows, reducing the risk of unintended harmful actions. The portability mechanism also promotes an open skill ecosystem where authors can write once and deploy anywhere, potentially accelerating the development of high-quality, reusable agent capabilities.

\textit{Potential Negative Impacts.}
Compile-time security analysis can produce false positives that block legitimate skills, potentially frustrating developers. The current interception rate (4.3\%) is low, but as rule sets expand, calibration will be important to maintain usability. Additionally, the existence of a compilation framework could create a false sense of security: \SkCC{} addresses format-level and pattern-level vulnerabilities but cannot guarantee semantic safety of skill logic, which ultimately depends on the skill author's intent and the agent's runtime behavior.

\textit{Mitigation Strategies.}
\SkCC{}'s diagnostic system provides non-blocking warnings for uncertain cases, allowing developers to review and override automated decisions. The tiered security classification enables graduated enforcement rather than binary accept/reject decisions. Future work on explainable security analysis could further improve developer trust and reduce false positive friction.

\section{How to Cite}

If you find \SkCC{} useful for your research, please consider citing:

\begin{lstlisting}[style=plaintext, numbers=none, frame=single, backgroundcolor=\color{black!3}]
@misc{ouyang2026skcc,
      title={SkCC: Portable and Secure Skill Compilation
             for Cross-Framework LLM Agents},
      author={Yipeng Ouyang and Yi Xiao and Yuhao Gu
              and Xianwei Zhang},
      year={2026},
      eprint={2605.03353},
      archivePrefix={arXiv},
      url={https://arxiv.org/abs/2605.03353},
}
\end{lstlisting}